\documentclass{llncs}
\usepackage{fullpage}
\usepackage{times}
\usepackage{color}
\usepackage{algorithmic}
\usepackage{algorithm}
\usepackage{graphicx}
\usepackage{subfig}
\usepackage{amsmath,amssymb}
\usepackage{tabularx}
\usepackage{setspace}
\usepackage[pdftex,bookmarks=true,colorlinks=true,linkcolor=black,citecolor=black,urlcolor=blue]{hyperref}
\usepackage{wrapfig}
\usepackage{tabularx}
\usepackage{amsfonts,amssymb}
\usepackage{mathrsfs}
\usepackage{marvosym}
\usepackage{pifont}

\newcommand{\bbn}{\mathbb{N}}

\newcommand{\calL}{{\cal L}}
\newcommand{\calP}{{\cal P}}
\newcommand{\calA}{{\cal A}}
\newcommand{\exec}{\mathit{exec}}
\newcommand{\true}{\ensuremath{\mathsf{true}}}

\newcommand{\pref}{\preccurlyeq}
\newcommand{\ef}{\ensuremath{E_{\varphi}}}
\newcommand{\efalgo}{\ensuremath{{E_{\varphi}^*}}}

\newcommand{\ptick}{\mathsf{ptick}}
\newcommand{\editI}{\mathsf{editI_{\varphi_I}}}
\newcommand{\editO}{\mathsf{editO_{\varphi}}}
\newcommand{\editIaut}{\mathsf{editI_{\calA_{\varphi_I}}}}
\newcommand{\editOaut}{\mathsf{editO_{\calA_\varphi}}}

\newcommand{\randEditIaut}{\mathsf{{nondet}\textendash editI_{\calA_{\varphi_{I}}}}}
\newcommand{\randEditOaut}{\mathsf{{nondet}\textendash editO_{\calA_\varphi}}}

\newcommand{\readInp}{\mathsf{read\_in\_chan}}
\newcommand{\readOut}{\mathsf{read\_out\_chan}}
\newcommand{\release}{\mathsf{release}}

\newcommand{\squishlist}{
 \begin{list}{$\bullet$}
  { \setlength{\itemsep}{0pt}
     \setlength{\parsep}{1pt}
     \setlength{\topsep}{1pt}
     \setlength{\partopsep}{0pt}
     \setlength{\leftmargin}{0.9em}
     \setlength{\labelwidth}{1.5em}
     \setlength{\labelsep}{0.4em} } }
\newcommand{\squishend}{
  \end{list}  }
\title{Runtime enforcement of reactive systems using synchronous enforcers\thanks{This work has been partially supported by the Academy of Finland, the U.S. National Science Foundation (awards \#1329759 and \#1139138), and the Deutsche Forschungsgemeinschaft (PRETSY2 project, award DFG HA 4407/6-2).}}
\author{Srinivas Pinisetty\inst{1}, Partha S Roop\inst{2}, Steven Smyth\inst{3}, Stavros Tripakis\inst{1,4},\\ Reinhard von Hanxleden\inst{3}}
\institute{Aalto University, Finland\quad \email{First.Last@aalto.fi}\and
University of Auckland, New Zealand\quad \email{p.roop@aucklanduni.ac.nz} \and
University of Kiel, Germany\quad \email{ssm,rvh@informatik.uni-kiel.de} \and
University of California, Berkeley, USA
}
\begin{document}
\maketitle
\pagestyle{plain}
\begin{abstract}
Synchronous programming is a paradigm of choice for the design of safety-critical reactive systems.
Runtime enforcement is a technique to ensure that the output of a black-box system satisfies some desired properties.
This paper deals with the problem of runtime enforcement in the context of synchronous programs.
We propose a framework where an enforcer monitors both the inputs and the outputs of a synchronous program and (minimally) edits erroneous inputs/outputs in order to guarantee that a given property holds.
We define enforceability conditions, develop an online enforcement algorithm, and prove its correctness.
We also report on an implementation of the algorithm on top of the KIELER framework for the SCCharts synchronous language.
Experimental results show that enforcement has minimal execution time overhead, which decreases proportionally with larger benchmarks.
\end{abstract}
\section{Introduction}
\label{sec:intro}
Runtime verification (RV)~\cite{LeuckerS08jlap,FalconeHR13} is an active area of research on methods that dynamically verify a set of desirable properties over an execution of a ``black-box'' system.
An alternative to such passive runtime analysis is runtime enforcement (RE)~\cite{enforceablesecpol,FalconeMFR11,RuntimeNonSafety,FMSD}.
In RE mechanisms, an \emph{enforcer} is synthesized to observe the executions of a black-box system to ensure that a set of desired properties are satisfied.
In the event of a violation, the enforcer performs certain evasive actions so as to prevent the violation.
The evasive actions might include blocking the execution~\cite{enforceablesecpol}, modifying input sequence
by suppressing and / or inserting actions~\cite{RuntimeNonSafety}, and buffering input actions until a future time when it could be forwarded~\cite{FalconeMFR11,FMSD}.
These enforcement mechanisms are not suitable for synchronous reactive systems since delaying the reaction or terminating the system is infeasible.
Considering this, there is recent interest in runtime enforcement of synchronous reactive systems~\cite{BloemKKW15}.

A synchronous reactive system is non-terminating and interacts continuously with the adjoining environment.
Hence, the system execution may be considered as a series of steps,
where in each step the system reads the inputs from the environment,
calls a \emph{reaction function} that computes the outputs for emission.
Synchronous programming languages~\cite{BenvenisteCEHLd03} are well suited for the design of synchronous reactive systems.
They use observers~\cite{HalbwachsLR94} to express safety properties, which are verified statically (using model checking).
There have also been limited attempts to use observers as runtime entities~\cite{RaymondNHW98}, for example for automatic test case generation.
More recently Rushby studies applications of observers for the expression of assumptions and axioms in addition to test case generation~\cite{Rushby14}.
However, there have been no studies on the bi-directional RE problem for synchronous reactive systems, which is the focus of the current paper.
\begin{figure}[!]
\begin{centering}
\includegraphics[scale=1]{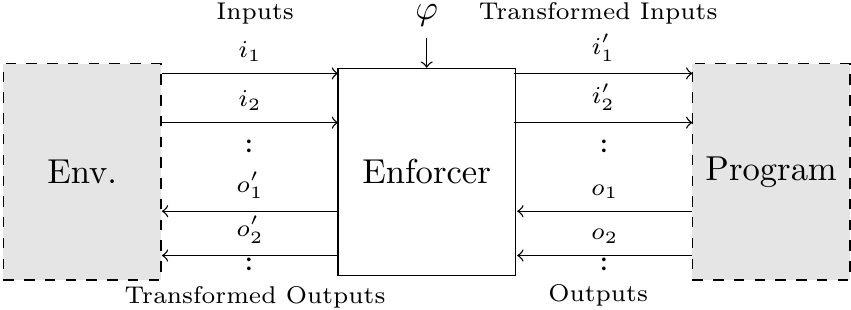}
\caption{Bi-directional enforcement for synchronous programs.}
\label{fig:intro-context}
\end{centering}
\end{figure}

We consider bi-directional RE of synchronous programs, and the general context is illustrated in Figure~\ref{fig:intro-context}.
Here, $\{ i_1, i_2, \cdots, i_n \}$ are inputs from the environment
to the enforcer, $\{ i'_1, i'_2, \cdots, i'_n \}$ are transformed inputs from the enforcer to the program,
$\{ o_1, o_2, \cdots, o_m \}$ are outputs of the program to the enforcer, and $\{ o'_1, o'_2, \cdots,  o'_m \}$ are transformed outputs from the enforcer to the environment.
RE for synchronous reactive systems is distinct from the existing RE mechanisms such as~\cite{FalconeMFR11,RuntimeNonSafety,FMSD,enforceablesecpol} since the enforcement mechanism for a synchronous reactive system cannot halt the system or delay events, and must react instantaneously when an error is observed.
Moreover, we consider bi-directional enforcement where the enforcer needs to consider the status of the environment and the program in order to enforce the policies.
The enforcer must respect the \emph{causality} aspects i.e. every reactive cycle must start with the environment,
where the status of the environment inputs must determine the reaction.
After the program has reacted, the generated outputs are emitted to the environment.
Considering this, the enforcer must act as an intermediary such that it first intercepts the inputs from the
environment to validate them relative to the policy and forward the
inputs to the program once the policy is satisfied. In the event of
any violation, the enforcer may suitably alter the inputs before forwarding to the program.
After the program has reacted to these
inputs, again the enforcer must ensure that either the policy is
satisfied and hence the outputs are forwarded unchanged to the
environment or a violation has happened that needs to be handled by
altering the outputs to prevent policy violation.

We study the problem of synthesizing an enforcer for any given safety property $\varphi$.
Similar to enforcement mechanisms in~\cite{enforceablesecpol,FalconeMFR11,RuntimeNonSafety,FMSD},
several constraints are required on how an enforcer transforms input-output words.
The enforcer cannot delay events, and cannot block execution,
but it is allowed to \emph{edit} an event when necessary (i.e., when the event that it receives as input leads to a violation).
The notions of \emph{soundness} and \emph{transparency} are similar to the existing
enforcement mechanisms~\cite{enforceablesecpol,FalconeMFR11,RuntimeNonSafety,FMSD},
 where soundness means that the output of the enforcer must satisfy property $\varphi$,
and transparency expresses that the enforcer should not modify events unnecessarily.
In the proposed framework, we also introduce
additional requirements called \emph{causality}, and \emph{instantaneity}. These constraints are developed specifically to respect synchronous execution, detailed in Section~\ref{sec:problemDef}.
\paragraph{Contributions.}
In this paper, we study and formally define, for the first time, the bi-directional enforcer synthesis problem for synchronous reactive systems (expressed as synchronous programs).
The main contributions of the paper are (1) We formally define the bi-directional enforcer synthesis problem and characterize the set of safety properties which can be enforced (Section~\ref{sec:problemDef}), (2) We develop an enforcement algorithm (Section~\ref{sec:algo}) and prove its correctness, (3) We report on an implementation of the algorithm on top of the KIELER framework for the SCCharts synchronous language (Section~\ref{sec:sccharts}), and (5) We evaluate the approach over a range of synchronous programs in the SCCharts language~\cite{vonHanxleden2014} to illustrate scalability and practicality (Section~\ref{sec:sccharts}).
\section{Preliminaries and Notation}
\label{sec:prelim}
A finite (resp. infinite) word over a finite alphabet $\Sigma$ is a finite sequence $\sigma = a_1\cdot a_2\cdots a_n$ (resp. infinite sequence $\sigma = a_1\cdot a_2\cdots$) of elements of $\Sigma$.
The set of finite (resp. infinite) words over $\Sigma$ is denoted by $\Sigma^*$ (resp. $\Sigma^\omega$).
The {\em length} of a finite word $\sigma$ is $n$ and is noted $|\sigma|$.
The empty word over $\Sigma$ is denoted by $\epsilon_\Sigma$, or $\epsilon$ when clear from the context.
The {\em concatenation} of two words $\sigma$ and $\sigma'$ is denoted as $\sigma\cdot \sigma'$.
A word $\sigma'$ is a {\em prefix} of a word $\sigma$, denoted as $\sigma' \pref \sigma$, whenever there exists a word $\sigma''$ such that $\sigma = \sigma'\cdot \sigma''$; $\sigma$ is said to be an \emph{extension} of $\sigma'$.

We consider a reactive system with a finite {ordered} sets of Boolean inputs $I= \{i_1, i_2,\cdots,i_n\}$ and Boolean outputs $O= \{o_1, o_2,\cdots,o_m\}$.
The input alphabet is $\Sigma_I=2^I$, and the output alphabet is $\Sigma_O=2^O$ and the input-output alphabet $\Sigma= \Sigma_I \times \Sigma_O$.
Each input (resp. output) event will be denoted as a bit-vector/complete monomial.
For example, let $I=\{A, B\}$.
Then, the input $\{A\} \in \Sigma_I$ is denoted as  $10$, while $\{B\} \in \Sigma_I$ is denoted as $01$ and  $\{A, B\} \in \Sigma_I$ is denoted as $11$.
A reaction (or input-output event) is of the form $(x_i, y_i)$, where $x_i \in \Sigma_I$ and $y_i \in \Sigma_O$.

Given an input-output word $\sigma= (x_1,y_1)\cdot(x_2,y_2)\cdots(x_n,y_n) \in \Sigma^*$, the input word obtained from $\sigma$ is $\sigma_I = x_1 \cdot x_2 \cdots x_n \in \Sigma_I$ which is the projection on inputs ignoring outputs.
Similarly, the output word obtained from $\sigma$ is $\sigma_O = y_1 \cdot y_2 \cdots y_n \in \Sigma_O$ is the projection on outputs.

An execution $\sigma$ of a synchronous program $\calP$ is an infinite sequence of input-output events $\sigma\in\Sigma^{\omega}$, and the \emph{behavior} of a synchronous program $\calP$ is denoted as $\exec(\calP)\subseteq \Sigma^{\omega}$.
The \emph{language} of $\calP$ is denoted by $\cal{L}(\cal{P})$ = $\{\sigma \in \Sigma^* | \exists \sigma' \in \exec(\calP) \wedge \sigma \pref \sigma'\}$ i.e. $\cal{L}(\cal{P})$ is the set of all finite prefixes of the sequences in $\exec(\calP)$.

A property $\varphi$ over $\Sigma$ defines a set $\calL(\varphi)\subseteq \Sigma^{*}$.
A program $\calP \models \varphi$ iff $\calL(\calP) \subseteq  \calL(\varphi)$.
Given a word $\sigma \in \Sigma^*$, $\sigma \models \varphi$ iff $\sigma\in\calL(\varphi)$.
A {property} $\varphi$ is {\em prefix-closed} if all prefixes of all words from
$\calL(\varphi)$ are also in $\calL(\varphi)$: $\calL(\varphi) = \{w\;|\;\exists w'\in\calL(\varphi): w\pref w'\}$.
In this paper, we consider prefix-closed properties.
Properties are formally expressed as safety automata that we define in the sequel.
\begin{definition}[Safety Automaton]
\label{def:SA}
A \emph{safety automaton} (SA) $\calA =(Q, q_0, q_v, \Sigma, \xrightarrow{})$ is a tuple, where $Q$ is the set of states, called \emph{locations}, $q_0 \in Q$ is an unique initial location, $q_v \in Q$ is a unique violating (non-accepting) location, $\Sigma=\Sigma_I\times\Sigma_O$ is the alphabet, and $\xrightarrow{} \subseteq Q \times \Sigma \times Q$ is the transition relation.
All the locations in $Q$ except $q_v$ (i.e., $Q \setminus \{q_v\}$) are accepting locations.
Location $q_v$ is a unique non-accepting (trap) location, and there are no transitions in $\xrightarrow{}$ from $q_v$ to a location in $Q \setminus \{q_v\}$.
Whenever there exists $(q, a, q') \in \xrightarrow{}$, we denote it as $q \xrightarrow{a} q'$.
Relation $\xrightarrow{}$ is extended to words $\sigma \in \Sigma^*$ by noting
$q \xrightarrow{\sigma . a} q'$ whenever there exists $q''$ such that $q
\xrightarrow{\sigma} q''$ and $q'' \xrightarrow{a} q'$.
A location $q\in Q$ is reachable from $q_0$ if there exists a word $\sigma \in \Sigma^*$ such that $q_0
\xrightarrow{\sigma} q$.
\end{definition}
An SA $\calA = (Q, q_0, q_v, \Sigma, \xrightarrow{})$ is \textit{deterministic} if $\forall
q \in Q, \forall a \in \Sigma, (q \xrightarrow{a} q' \land q \xrightarrow{a}
q'') \implies (q' = q'')$.
$\mathcal{A}$ is \textit{complete} if $\forall q \in Q, \forall a \in \Sigma, \exists q' \in Q, q \xrightarrow{a} q'$.
A word $\sigma$ is \textit{accepted} by $\mathcal{A}$ if there exists $q \in Q \setminus \{q_v\}$ such that $q_0
\xrightarrow{\sigma} q$.
The set of all words accepted by $\mathcal{A}$ is denoted as $\mathcal{L}(\mathcal{A})$.
\begin{remark}
In the rest of this paper, $\varphi$ is a safety property defined as deterministic and complete SA $\calA_\varphi = (Q, q_0, q_v, \Sigma, \xrightarrow{})$.
If the user provides an non-deterministic or incomplete automaton, we determinize and complete it first.
We also consider that $Q$ does not contain any (redundant) locations that are unreachable from $q_0$.
\end{remark}
Due to the causality requirement, the enforcer has to first transform inputs from the environment in each step according to property $\varphi$ defined as SA $\calA_\varphi$.
We thus need to consider the input property that we obtain from $\calA_\varphi$ by projecting on inputs.
\begin{definition}[Input safety automaton $\calA_{\varphi_I}$]
\label{def:inp:prop:proj:def}
Given $\varphi\subseteq\Sigma^*$, defined as SA $\calA_{\varphi}=(Q, q_0, q_v, \Sigma, \rightarrow)$,
input SA $\calA_{\varphi_I}=(Q, q_0, q_v, \Sigma_I, \rightarrow_I)$ is obtained from $\calA_{\varphi}$ by ignoring outputs on the transitions, i.e.,
for every transition  $q \xrightarrow{(x,y)} q' \in \rightarrow$ where $(x,y) \in \Sigma$, there is a transition  $q \xrightarrow{x} q' \in \rightarrow_I$, where $x \in \Sigma_I$.
$\calL(\calA_{\varphi_I})$ is denoted as $\varphi_I \subseteq \Sigma_I^*$.
\end{definition}
\begin{figure}[htb]
\centering
\subfloat[SA $\calA_{S_1}$. \label{fig:prop1}]{
\includegraphics[scale=1]{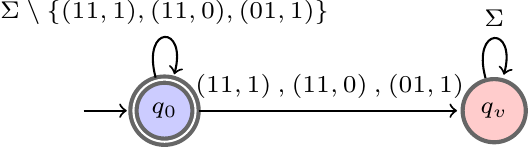}
}
\hspace{2.0em}
\subfloat[Input SA obtained from $\calA_{S_1}$. \label{fig:prop1Inp}]{
\includegraphics[scale=1]{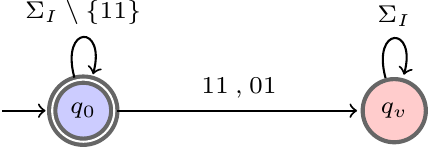}
}
\caption{SA (left), and its input SA (right).}
\label{fig:prop:inpProj}
\end{figure}
%
\begin{example}[Example property defined as SA and its input SA]
\label{eg:prop}
Let $I= \{A,B\}$ and $O = \{R\}$.
Consider the following property: \textit{$S_1$: ``A and B cannot happen simultaneously, and also B and R cannot happen simultaneously''}.
The safety automaton in Figure~\ref{fig:prop1} defines property $S_1$.
Figure~\ref{fig:prop1Inp} presents the input SA for the SA in Figure~\ref{fig:prop1} defining property $S_1$.
Though the SA $\calA_\varphi$ is deterministic, the input SA $\calA_{\varphi_I}$ might be non-deterministic as is the case in Figure~\ref{fig:prop1Inp}.
\end{example}
\begin{lemma}
\label{lem:inputProp}
Let $\calA_{\varphi_I}=(Q, q_0, q_v, \Sigma_I, \rightarrow_I)$ be the input automaton obtained from $\calA_{\varphi}=(Q, q_0, q_v, \Sigma, \rightarrow)$. We have the following properties:
\squishlist
\item[1] $\forall (x,y) \in \Sigma, \forall q, q'\in Q: q\xrightarrow{(x,y)}q' \implies q \xrightarrow{x}_I q'$.
\item[2] $\forall x \in \Sigma_I, \forall q, q'\in Q: q \xrightarrow{x}_I q' \implies \exists y\in\Sigma_O: q\xrightarrow{(x,y)}q'$.
\squishend
\end{lemma}
Intuitively, property 1 of Lemma~\ref{lem:inputProp} states that if there is a transition from state $q\in Q$ to state $q' \in Q$ upon input-output event $(x,y)\in \Sigma$ in the automaton $\calA_\varphi$, then there is also a transition from state $q$ to state $q'$ in the input automaton $\calA_{\varphi_I}$ upon the input event $x\in\Sigma_I$.
Property 2 of Lemma~\ref{lem:inputProp} states that if there is a transition from state $q\in Q$ to state $q' \in Q$ upon input event $x\in\Sigma_I$, then there certainly exists an output event $y\in\Sigma_O$ s.t. there is a transition from state $q$ to state $q'$ upon event $(x,y)$ in the automaton $\calA_\varphi$. Lemma~\ref{lem:inputProp} immediately follows from Definitions~\ref{def:SA} and~\ref{def:inp:prop:proj:def}.
\subsubsection{Edit Functions}
\label{sec:prelim:re}
Consider property $\varphi\subseteq\Sigma^*$ defined as SA $\calA_{\varphi}=(Q, q_0, q_v, \Sigma, \rightarrow)$, and its input SA $\calA_{\varphi_I}=(Q, q_0, q_v, \Sigma_I, \rightarrow_I)$ obtained from $\calA_{\varphi}$ by projecting on inputs.
We introduce $\editI$ (resp. $\editO$), that the enforcer uses for editing input (resp. output) events (when necessary), according to input property $\varphi_I$ (resp. property $\varphi$).
\begin{itemize}
\item {{\boldmath$\editI(\sigma_I)$}}: Given $\sigma_I\in\Sigma_I^*$, $\editI(\sigma_I)$ is the set of input events $x$ in $\Sigma_I$ such that the word obtained by extending $\sigma_I$ with $x$ satisfies property $\varphi_I$. Formally,
    \[\editI(\sigma_I) = \{ x\in \Sigma_I: \sigma_I \cdot x \models \varphi_I \}.\]
Considering the SA $\calA_{\varphi_I}=(Q, q_0, q_v, \Sigma_I, \rightarrow_I)$,
the set of events in $\Sigma_I$ that allow to reach a state in $Q\setminus \{q_v\}$ from a state $q\in Q\setminus \{q_v\}$ is defined as:
\[\editIaut(q) = \{x\in \Sigma_I: q \xrightarrow{x}_I q' \wedge q' \neq q_v \}. \]
For example, consider the SA in Figure~\ref{fig:prop1Inp} obtained from the SA in Figure~\ref{fig:prop1} by ignoring outputs.
Let $\sigma= (10,0)\cdot(01,1)$, and thus $\sigma_I= 10\cdot01$.
Then, $\editI(\sigma_I) = \Sigma_I\setminus\{11\}$.
Also, $q_0 \xrightarrow{10\cdot01}_I q_0$, and $\editIaut(q_0) = \Sigma_I\setminus\{11\}$.

\item{{\boldmath$\randEditIaut(q)$}}: If $\editIaut(q)$ is non-empty, then $\randEditIaut(q)$ returns an element (chosen randomly) from $\editIaut(q)$, and is undefined if $\editIaut(q)$ is empty.
\item {\boldmath$\editO(\sigma, x)$}:~~Given an input-output word $\sigma\in\Sigma^*$ and an input event $x\in\Sigma_I$, $\editO(\sigma, x)$ is the set of output events $y$ in $\Sigma_O$ s.t. the input-output word obtained by extending $\sigma$ with $(x,y)$ satisfies property $\varphi$. Formally,
\[\editO(\sigma,x) = \{y \in \Sigma_O: \sigma \cdot (x,y) \models \varphi \}.\]
Considering the automaton $\calA_{\varphi}=(Q, q_0, q_v, \Sigma, \rightarrow)$ defining property $\varphi$, and an input event $x\in\Sigma_I$,
the set of output events $y$ in $\Sigma_O$ that allow to reach a state in $Q\setminus \{q_v\}$ from a state $q\in Q\setminus \{q_v\}$ with $(x,y)$ is defined as:
\[\editOaut(q,x) = \{y \in \Sigma_O: q \xrightarrow{(x,y)} q' \wedge q' \neq q_v \}. \]
\noindent
For example, consider property $S_1$ defined by the automaton in Figure~\ref{fig:prop1}.
We have $\editOaut(q_0, 01) = \{0\}$.

\item{{\boldmath$\randEditOaut(q, x)$}}: If $\editOaut(q,x)$ is non-empty, then $\randEditOaut(q, x)$ returns an element (chosen randomly) from $\editOaut(q,x)$, and is undefined if $\editOaut(q,x)$ is empty.
\end{itemize}
\section{Problem Definition}
\label{sec:problemDef}
%
In this section, we formalize the RE problem for synchronous programs.
In the setting we consider, as illustrated in Figure~\ref{fig:intro-context}, an enforcer monitors and corrects both inputs and outputs of a synchronous program according to a given safety property $\varphi\subseteq\Sigma^*$.
We assume that the ``black-box'' synchronous program may be invoked through a special function call called $\ptick$, which is invoked exactly once
 during each reaction / synchronous step.
Formally, $\ptick$ is a function from $\Sigma_I$  to  $\Sigma_O$ that takes a bit vector $x \in \Sigma_I$ and returns a bit vector $y \in \Sigma_O$.

An enforcer for a property $\varphi$ can only edit an input-output event when necessary, and it cannot block, delay or suppress events.
Let us recall the two functions $\editI$ and $\editO$ that were introduced in Section~\ref{sec:prelim} that the enforcer for $\varphi$ uses to edit the current input (respectively output) event according to the property $\varphi$.
At an abstract level, an enforcer can be seen as a function that transforms input-output words.
An enforcement function for a given property $\varphi$ takes as input an input-output word over $\Sigma$ and outputs an input-output word over $\Sigma$ that belongs to $\varphi$.
%
\begin{definition}[Enforcer for $\varphi$]
\label{def-E-func-constraints}
Given property $\varphi\subseteq\Sigma^*$, an {\em enforcer} for $\varphi$ is a function $\ef: \Sigma^*\rightarrow \Sigma^*$ satisfying the following constraints:

\noindent
{\bf Soundness}
\begin{equation}
   \tag{\bf Snd}\label{eq:snd}
   \forall \sigma \in \Sigma^*: \ef(\sigma) \models \varphi.
\end{equation}
{\bf Monotonicity}
\begin{equation}
\tag{\bf Mono}\label{eq:mono}
\forall \sigma, \sigma' \in \Sigma^*: \sigma\pref \sigma' \Rightarrow \ef(\sigma) \pref \ef(\sigma').
\end{equation}
{\bf Instantaneity}
\begin{equation}
\tag{\bf Inst}\label{eq:inst}
\forall \sigma \in \Sigma^*: |\sigma| =  |\ef(\sigma)|.
\end{equation}
{\bf Transparency}
\begin{equation}
   \tag{\bf Tr}\label{eq:tr}
   \begin{array}{ll}
   \forall \sigma\in \Sigma^*, \forall x \in \Sigma_I,\forall y \in \Sigma_O:\\
             ~~~~~\ef(\sigma)\cdot(x,y)\models \varphi \implies \ef(\sigma\cdot(x,y)) = \ef(\sigma)\cdot(x,y).
   \end{array}
   \end{equation}
{\bf Causality}
\begin{equation}
   \tag{\bf Cau}\label{eq:ca}
  \begin{array}{ll}
   \forall \sigma \in \Sigma^*, \forall x \in \Sigma_I,\forall y \in \Sigma_O,\exists x' \in \editI(\ef(\sigma)_I), \\
 ~~~~~~~~~~~~~~~\exists y' \in \editO(\ef(\sigma), x'): \ef(\sigma\cdot(x,y))= \ef(\sigma)\cdot(x',y').
    \end{array}
\end{equation}
\end{definition}
The input-output sequence released as output by the enforcer upon reading the input-output sequence $\sigma$ is $\ef(\sigma)$, and $\ef(\sigma)_I \in \Sigma_I^*$ is the projection on the inputs.
Note, $\editI(\ef(\sigma)_I)$ returns a set of input events in $\Sigma_I$, s.t. $\ef(\sigma)_I$ (which is the projection of input-output word $\ef(\sigma)$ to the input alphabet) followed by any event in $\editI(\ef(\sigma)_I)$ satisfies $\varphi_I$.
$\editO(\ef(\sigma), x')$ returns a set of output events in $\Sigma_O$, s.t. for any event $y$ in $\editO(\ef(\sigma), x')$, $\ef(\sigma)\cdot(x',y)$ satisfies $\varphi$.
\begin{itemize}
\item Soundness (\ref{eq:snd}) means that for any word $\sigma\in\Sigma^*$, the output of the enforcer $\ef(\sigma)$ must satisfy $\varphi$.
\item Monotonicity (\ref{eq:mono}) expresses that the output of the enforcer for an extended word $\sigma'$ of a word $\sigma$, extends the output produced by the enforcer for $\sigma$.
The monotonicity constraint means that the enforcer cannot undo what is already released as output.
\item Instantainety (\ref{eq:inst}) expresses that for any given input-output word $\sigma$ as input to the enforcer, the output of the enforcer $\ef(\sigma)$ should contain exactly the same number of events that are in $\sigma$ (i.e., $\ef$ is length-preserving).
This means that the enforcer cannot delay, insert and suppress events.
Whenever the enforcer receives a new event, it has to react instantaneously and has to produce an output event immediately.
\item
Transparency (\ref{eq:tr}) expresses that for any given word $\sigma$ and any event $(x,y)$, if the output of the enforcer for $\sigma$ (i.e., $\ef(\sigma)$) followed by the event $(x,y)$ satisfies the property $\varphi$ (i.e., $\ef(\sigma)\cdot(x,y) \models \varphi$), then the output that the enforcer produces for input $\sigma\cdot(x,y)$ will be $\ef(\sigma)\cdot(x,y)$.
This means that the enforcer makes no change when no change is needed in order to satisfy the property $\varphi$.
\item Causality (\ref{eq:ca}) expresses that for every input-output event $(x,y)$ the enforcer produces input-output event $(x',y')$ where the enforcer first processes the input part $x$, to produce the transformed input $x'$ according to property $\varphi$ using $\editI$.
The enforcer later reads and transforms output $y\in\Sigma_O$ which is the output of the program after invoking function $\ptick$ with the transformed input $x'$, to produce the transformed output $y'$ using $\editO$.
\end{itemize}
\begin{remark}
\label{rem:edt}
Let $\ef(\sigma)$ be the input-output sequence released as output by the enforcer for $\varphi$ after reading input-output sequence $\sigma \in \Sigma^*$.
Upon reading a new event $(x,y)$, if what has been already computed as output by the enforcer $\ef(\sigma)$ followed by $(x,y)$ does not allow to satisfy the property $\varphi$, then the enforcer edits $(x,y)$ using functions $\editI$ and $\editO$.
When the current event $(x,y)$ has to be edited, note that there may be several possible solutions.
For example, consider the property $S_1$ introduced in Example~\ref{eg:prop}.
Let $\sigma = (10,1)\cdot(01,0)$, and the output of the enforcer after processing $\sigma$ will be $\ef(\sigma) = (10,1)\cdot(01,0)$.
Let the new event be $(11,0)$, and $\ef(\sigma)\cdot(11,0)\not\models\varphi$, and the enforcer has to edit the new event $(11,0)$.
Note that $\ef(\sigma)_I= 10\cdot01$, and $\editI(10\cdot01) = \{00,01,10\}$ and the enforcer can choose any element from $\editI(10\cdot01)$ as the transformed input.
\end{remark}
\begin{remark}[Enforcing bi-directional properties]
By considering two uni-directional enforcers, where one enforcer checks and transforms inputs from the environment to the program and another enforcer checks and transforms outputs from the program to the environment, bi-directional properties cannot be enforced.
For example, bi-directional properties such as the property $S_1$ introduced in Example~\ref{eg:prop} cannot be enforced using two uni-directional enforcers.
\end{remark}
\begin{remark}[When the input word provided to the enforcer satisfies $\varphi$]
Constraint~(\ref{eq:tr'}) expresses that when any input-output word $\sigma\in\Sigma^*$ provided as input to the enforcer satisfies the property $\varphi$, then the enforcer will not edit any event and will output $\sigma$ (i.e., $\ef(\sigma) = \sigma$).
\begin{equation}
   \tag{\bf Tr'}\label{eq:tr'}
   \begin{array}{ll}
   \forall \sigma\in \Sigma^*:\ef(\sigma)\models \varphi \implies \ef(\sigma) = \sigma.
   \end{array}
\end{equation}
\end{remark}
\begin{lemma}
\label{lem:tr:tr'}
(\ref{eq:tr}) $\Rightarrow$ (\ref{eq:tr'}).

Lemma~\ref{lem:tr:tr'} shows that~(\ref{eq:tr'}) is a consequence of constraint~(\ref{eq:tr}).
For any $\varphi$, for any $\sigma\in\Sigma^*$, proof of this lemma is straightforward using induction on $\sigma$.
\end{lemma}
%
\vspace*{-\baselineskip}
\begin{table}[htb]
\centering
\begin{tabular}{|c|c|c|c|}
\hline
\multicolumn{1}{|c|}{$\sigma$} & $E_\varphi(\sigma)$ & \ref{eq:tr} &  \ref{eq:tr'} \\
\hline
$(10,1)$ & $(10,1)$ & \ding{51} & \ding{51}\\
\hline
$(10,1)\cdot (11,1)$ & $(10,1)\cdot \textbf{(10,1)}$ & \ding{51} & \ding{51}\\
\hline
$(10,1)\cdot (11,1)\cdot(01,0)$ & $(10,1)\cdot \textbf{(10,1)}\cdot\textbf{(10,0)}$ & \ding{55} & \ding{51}\\
\hline
$(10,1)\cdot (11,1)\cdot(01,0)$ & $(10,1)\cdot \textbf{(10,1)}\cdot(01,0)$ & \ding{51} & \ding{51}\\
\hline
\end{tabular}
\caption{Example: (\ref{eq:tr}) Vs. (\ref{eq:tr'})}
\label{tableExampleTR}
\end{table}
\begin{example}[(\ref{eq:tr}) is stronger than (\ref{eq:tr'})]
Via this example, we illustrate that constraint~(\ref{eq:tr}) is stronger than the alternative transparency constraint~(\ref{eq:tr'}).
Let us consider the property $S_1$ introduced in Example~\ref{eg:prop}.
In Table~\ref{tableExampleTR}, first column denoted using $\sigma$ shows input-output words, and the second column denoted using $\ef(\sigma)$ shows the output of the enforcer for $\sigma$, and the next two columns indicate whether $\ef(\sigma)$ satisfies constraints (\ref{eq:tr}) and (\ref{eq:tr'}) respectively.
We can see that there are situations where (\ref{eq:tr'}) holds and (\ref{eq:tr}) does not hold.
When the enforcer reads the third event $(01,0)$, if it edits this event to $(10,0)$, then constraint~(\ref{eq:tr'}) holds, and constraint~(\ref{eq:tr}) does not hold since $\ef((10,1)\cdot (11,1))$ followed by the new event read $(01,0)$ satisfies the property $S_1$, and it should not be edited by the enforcer according to constraint (\ref{eq:tr}).
\end{example}
\begin{definition}[Enforceability]
Let $\varphi\subseteq\Sigma^*$ be a property. We say that
$\varphi$ is {\em enforceable} iff an enforcer $\ef$ for $\varphi$ exists according to Definition~\ref{def-E-func-constraints}.
\end{definition}
Not all properties are enforceable,
even if we restrict ourselves to prefix-closed safety properties,
as the following example shows.
%
\begin{figure}[htb]
\centering
\includegraphics[scale=1]{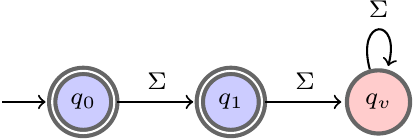}
\caption{A non-enforceable safety property.}
 \label{fig:prop1:nonEnf}
\end{figure}
\begin{example}[Non-enforceable safety property]
\label{eg:nonEnf}
We illustrate that not all prefix-closed safety properties are enforceable according to Definition~\ref{def-E-func-constraints}.
Consider the automaton in Figure~\ref{fig:prop1:nonEnf} defining the property $\varphi$ that we want to enforce, with $I = \{ A \}$, $O = \{ B \}$ and $\Sigma = \Sigma_I \times \Sigma_O$.
Let the input-output sequence provided as input to the enforcer be $\sigma=(1,1)\cdot(1,0)$.
When the enforcer reads the first event $(1,1)$, it can output $(1,1)$ (since every event in $\Sigma$ from $q_0$ leads to a non violating state $q_1$).
Note that from $q_1$, every event in $\Sigma$ only leads to violating state $q_v$.
Thus, when the second event $(1,0)$ is read, every possible editing of this event will only lead to violation of the property.
Upon reading the second event $(1,0)$, releasing any event in $\Sigma$ as output will violate soundness, and if no event is released as output, then the instantianety constraint will be violated.
\end{example}
\begin{theorem}[Condition for enforceability]
\label{rem:nonEnf}
Consider a property $\varphi$ defined as SA $\calA_\varphi=(Q, q_0, q_v, \Sigma, \rightarrow)$.
Property $\varphi$ is enforceable iff the following condition holds:
\begin{equation}
\label{suff_cond_enforceability}
\tag{\bf EnfCo}
\forall q \in Q, q \neq q_v \implies \exists (x,y) \in \Sigma: q \xrightarrow{(x,y)} q' \wedge q'\neq q_v
\end{equation}
\end{theorem}
%
Proof of Theorem~\ref{rem:nonEnf} is given in Appendix~\ref{sec:proofs}, page~\pageref{proof:rem:nonEnf}.
Note that given any property $\varphi$ defined as SA $\calA_\varphi=(Q, q_0, q_v, \Sigma, \rightarrow)$, it is straightforward to test whether $\calA_\varphi$ satisfies condition~(\ref{suff_cond_enforceability}).
\begin{figure}[htb]
\centering
\subfloat[A non-enforceable property that can be transformed into an enforceable property. \label{fig:nonenf:prop2}]{
\includegraphics[scale=0.9]{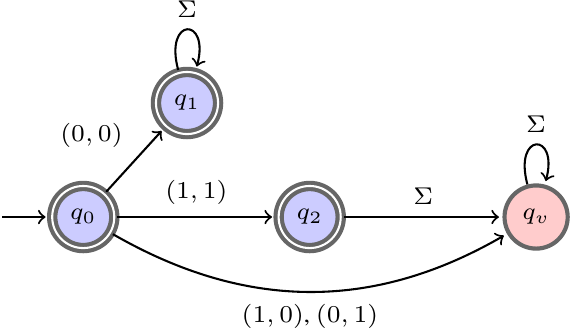}
}
\hspace{4em}
\subfloat[Enforceable property obtained after transformation. \label{fig:nonenf:prop2:trans}]{
\includegraphics[scale=0.9]{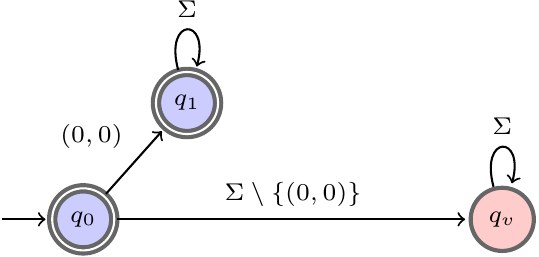}
}
\caption{A non-enforceable property transformed into an enforceable property.}
\label{trans:nonEnf:enf}
\end{figure}
\begin{remark}[Transforming a non-enforceable property into an enforceable property]
\label{rem:transfo:nonEnf}
Some non-enforceable properties can be made enforceable by a transformation that excludes some behaviors from the property.
We illustrate this with an example.
Consider the property defined by the automaton in Figure~\ref{fig:nonenf:prop2}.
This property is not enforceable for the following reason.
Suppose that the first input-output event that the enforcer receives is $(1,1)$.
Since there is a transition from $q_0$ to $q_2$ upon $(1,1)$, the enforcer will take this transition (according to transparency constraint).
Then, whatever may be the second event that the enforcer receives, note that $\editI$ and $\editO$ will be empty, and there is no way to correct the event and avoid reaching $q_v$.
However, we can transform this property into an enforceable property by
excluding all the paths/behaviours that are problematic.
In particular, we can remove state $q_2$ from the automaton of
Figure~\ref{fig:nonenf:prop2} and redirect the transition labeled $(1,1)$
from $q_0$ to $q_v$ instead.
This has the effect of removing the word $(1,1)$ from the language accepted by this automaton.
The resulting automaton (shown in Figure~\ref{fig:nonenf:prop2:trans}) that we obtain satisfies the condition for enforceability~(\ref{suff_cond_enforceability})
and therefore the resulting new property is enforceable.
Note that transforming a non-enforceable property to an enforceable one is not always possible.
For instance, the non-enforceable property of Figure~\ref{fig:prop1:nonEnf} cannot be transformed to an enforceable property.
\end{remark}
\paragraph{Transformation of non-enforceable properties.}
If a given safety property $\varphi$ defined as SA $\calA_\varphi=(Q, q_0, q_v, \Sigma, \xrightarrow{})$ does not satisfy the condition for enforceability~(\ref{suff_cond_enforceability}), then we can apply the following transformation process to check whether $\calA_\varphi$ can be transformed in to an enforceable property (by discarding some states in $ Q\setminus\{q_0\}$ in the automaton $\calA_\varphi$).
We discuss the algorithm for transformation briefly.
\begin{itemize}
\item For every state $q\in Q\setminus\{q_v\}$ if $\forall (x,y) \in \Sigma, q \xrightarrow{(x,y)} q_v$, then merge $q$ with $q_v$ ($q$ is removed from the set of states $Q$ and all the incoming transitions to $q$ go to $q_v$ instead).
\item The transformation continues until one of the following two conditions hold:
\begin{itemize}
\item only two states $q_0$ and $q_v$ remain in $Q$, i.e., $Q = \{q_0, q_v\}$ such that $\forall (x,y) \in \Sigma, q_0 \xrightarrow{(x,y)} q_v$. In this case, the algorithm returns that $\calA_\varphi$ cannot be transformed into an enforceable property.
\item $Q\setminus\{q_0,q_v\}$ is non-empty, and there is no state in $Q\setminus\{q_0,q_v\}$, that has all its outgoing transitions to $q_v$. In this case, the algorithm returns the resulting transformed automaton which is an enforceable property. Let $\mathsf{sub}(\calA_\varphi)$ be the transformed automaton. Note that $\calL(\mathsf{sub}(\calA_\varphi))\subseteq \calL(\calA_\varphi)$.
\end{itemize}
\end{itemize}

The algorithm for transformation of non-enforceable properties is discussed in detail in Appendix~\ref{ap:trans:nonEnf}.
\section{Algorithm}
\label{sec:algo}
In this section, we provide an algorithm for implementing the bi-directional synchronous enforcement problem defined in Section~\ref{sec:problemDef}.
Let the SA $\calA_{\varphi}=(Q, q_0, q_v, \Sigma, \rightarrow)$ define the property $\varphi$ that we want to enforce.
SA $\calA_{\varphi_I}=(Q, q_{0}, q_{v}, \Sigma_I, \rightarrow_I)$ is obtained from $\calA_{\varphi}$ by projecting on inputs (see section~\ref{sec:prelim:re}).
%
\begin{algorithm}[htb]
\caption{$\mathsf{Enforcer}$}
\label{algo:enf}
{
\begin{algorithmic}[1]
   \STATE $t \gets 0$
    \STATE $q \gets q_{0}$
    \WHILE {$\true$}
        \STATE $x_t \gets \readInp()$
        \label{algoEi-readIn}
        \label{algoEi-begin}
        \IF {$\exists q'\in Q: q \xrightarrow{x_t}_I q' \wedge q' \neq q_{v}$}
        \label{algoEi-test}
            \STATE $x'_t \gets x_t$
        \label{algoEi-noEdit}
        \ELSE
            \STATE $x'_t \gets \randEditIaut(q)$
            \label{algoEi-edit}
            \label{algoEi-Edit}
        \ENDIF
        \label{algoEi-end}
        \STATE $\mathsf{\ptick(x'_t)}$
        \STATE $y_t \gets \readOut()$
        \label{algoEi-readOut}
         \label{algoEo-begin}
        \IF {$\exists q'\in Q: q\xrightarrow{(x'_t,y_t)} q' \wedge q' \neq q_{v}$}
        \label{algoEo-test}
            \STATE $y'_t \gets y_t$
        \label{algoEo-noEdit}
        \ELSE
             \STATE $y'_t \gets \randEditOaut(q, x'_t)$
             \label{algoEo-edit}
        \label{algoEo-Edit}
        \ENDIF
        \label{algoEo-end}
        \STATE $\release((x'_t, y'_t))$
        \label{algoEo-release}
        \STATE $q \gets q'$ ~~~~ {\footnotesize{where $q\xrightarrow{(x'_t,y'_t)} q' \wedge q'\neq q_v$}}
        \label{algo-stateUpdate}
        \STATE $t \gets t+1$
        \ENDWHILE
\end{algorithmic}
}
\end{algorithm}
%

We provide an online algorithm that requires automata $A_{\varphi}$ and $A_{\varphi_I}$ as input.
Algorithm~\ref{algo:enf} is an infinite loop, and an iteration of the algorithm is triggered at every time step.
We  adapt the \emph{reactive interface} that is used for linking the program to its adjoining environment by following the structure of the interface  described in~\cite{andre01}. We extend the interface by including the enforcer as an intermediary between the synchronous program and its adjoining environment.

In the algorithm shown below, $t$ keeps track of the time-step (\emph{tick}), initialized with 0.
$q$ keeps track of the current state of both the automata $\calA_\varphi$ and $\calA_{\varphi_I}$.
Recall that the automaton $\calA_{\varphi_I}$ that we obtain from the automaton $\calA_{\varphi}$ by projecting on inputs (see Section~\ref{sec:prelim:re}) have identical structure, and the only difference is that the outputs are ignored on the transitions in the automaton $\calA_{\varphi_I}$.
Note that at the beginning of each iteration of the algorithm, the current states of both the automata $\calA_\varphi$ and $\calA_{\varphi_I}$ are the same (where both are initialized with $q_0$).
At $t$, if $\mathsf{EOut}\in \Sigma^*$ is the input-output sequence obtained by concatenating all the events released as output by the enforcer until time $t$, then $q$ corresponds to the state that we reach in the automaton $\calA_\varphi$ upon reading $\mathsf{EOut}$.
Similarly, if $\mathsf{EOut_{I}} \in \Sigma^*_I$ is the sequence obtained by projecting on $x_i's$ from $\mathsf{EOut}$, $q$ also corresponds to the state that we reach in the automaton $\calA_{\varphi_I}$ upon reading $\mathsf{EOut_{I}}$.

Functions $\readInp$ (resp. $\readOut$) are functions corresponding to reading input (resp. output) channels, and function $\ptick$ corresponds to invoking the synchronous program.
Function $\release$ takes an input-output event, and releases it as output of the enforcer.

Each iteration of the algorithm proceeds as follows:
first all the input channels are read using function $\readInp$ and the input event is assigned to $x_t$.
Then the algorithm tests whether there exists a transition in $\rightarrow_I$ from the current state $q$ upon $x_t$ to an accepting state in $\calA_{\varphi_I}$.
In case if this test succeeds, then it is not necessary to edit the input event $x_t$, and the transformed input $x'_t$ is assigned $x_t$.
Otherwise, $x'_t$ is assigned with the output of $\randEditIaut(q)$.
Let us recall that $\randEditIaut(q)$ returns an input event that leads to an accepting state in $\calA_{\varphi_I}$ from $q$.

After transforming the input $x_t$ according to $\calA_{\varphi_I}$, the program is invoked with the transformed input $x'_t$ using function $\ptick$.
Afterwards, all the output channels are read using function $\readOut$ and the output event is assigned to $y_t$.
Then the algorithm tests whether there exists a transition in $\rightarrow$ from the current state $q$ upon $(x'_t, y_t)$ to an accepting state in $\calA_{\varphi}$.
In case if this test succeeds, then it is not necessary to edit the output event $y_t$, and the transformed output $y'_t$ is assigned $y_t$.
Otherwise, $y'_t$ is assigned with the output of $\randEditOaut(q, x'_t)$.
Note that $\randEditOaut(q, x'_t)$ returns an output event $y'_t$ such that $(x'_t, y'_t)$ leads to an accepting state in $\calA_{\varphi}$ from $q$.

Before proceeding with the next iteration, current state $q$ is updated to $q'$ which is the state reached upon $(x'_t, y'_t)$ from state $q$ in the automaton $\calA_\varphi$, and the time-step $t$ is incremented.
Note that if there exists a transition $q \xrightarrow{(x'_t, y'_t)} q'$ in the SA $\calA_\varphi$, then there also exists a transition $q\xrightarrow{x'_t}_I q'$ in the SA $\calA_{\varphi_I}$.
The current states of both the SA are always synchronized and the same at the beginning of each iteration of the algorithm.
\begin{definition}[$\efalgo$]
\label{def-algo-ef}
Consider an enforceable safety property $\varphi$.
We define the function $\efalgo:\Sigma^*\to\Sigma^*$, where $\Sigma=\Sigma_I\times\Sigma_O$, as follows.
Let $\sigma = (x_1, y_1) \cdots (x_k,y_k)\in\Sigma^*$ be a word received by
Algorithm~\ref{algo:enf}. Then we let
$\efalgo(\sigma)=(x_1', y_1') \cdots (x_k',y_k')$, where $(x_t',y_t')$ is the
pair of events output by Algorithm~\ref{algo:enf} in Step~\ref{algoEo-release},
for $t=1,...,k$.
\end{definition}
%
\begin{theorem}[Correctness of the enforcement algorithm]
\label{prop-constraints-algo}
Given any safety property $\varphi$ defined as SA $\calA_\varphi$ that satisfies condition~(\ref{suff_cond_enforceability}),
the function $\efalgo$ defined above is an enforcer for $\varphi$, that is,
it satisfies (\ref{eq:snd}), (\ref{eq:tr}), (\ref{eq:mono}), (\ref{eq:inst}), and (\ref{eq:ca}) constraints of Definition~\ref{def-E-func-constraints}.
\end{theorem}
Proof of Theorem~\ref{prop-constraints-algo} is given in Appendix~\ref{sec:proofs}, page~\pageref{proof-prop-constraints-algo}.
%
\begin{remark}[Determinism of the enforcer]
Since we consider synchronous programs, the enforcer should be deterministic.
Regarding determinism, note that though $\calA_\varphi$ is deterministic, the enforcer $\efalgo$ may be non-deterministic, because when the received input $x$ (resp. output $y$) does not lead to an accepting state from the current state $q$ in $\calA_{\varphi_I}$, (resp. $\calA_\varphi$), it is edited in step~\ref{algoEi-Edit} (resp. step~\ref{algoEo-Edit}) of the algorithm.
Note that $\editIaut(q)$ (resp. $\editOaut(q,x)$ where $x\in \editIaut(q)$), may contain more than one element as illustrated via an example in Remark~\ref{rem:edt}, and $\randEditIaut$ (resp. $\randEditOaut$) will choose one element from the set $\editIaut$ (resp. $\editOaut$).
However, it is straightforward to make the behavior deterministic by computing $\editIaut(q)$ off-line for all $q\in Q\setminus\{q_v\}$, and selecting one element randomly from $\editIaut(q)$ and remembering the selection for each $q$ by storing in a table with size $|Q|$.
Thus, whenever in some state $q$ and when the input read $x$ does not lead to an accepting state from $q$ (i.e, the condition tested in line~\ref{algoEi-test} evaluates to false), in step~\ref{algoEi-Edit} we check the element corresponding to the state $q$ from the table and assign it to $x'$.
Similarly, for event $q\in Q\setminus\{q_v\}$, and for all $x\in\editIaut(q)$, we can compute $\editOaut(q,x)$ off-line, select one element randomly and store the selection in a table with size $|Q\times\Sigma_I|$.
Thus, whenever in some state $q$, when $(x',y)$ (where $x'$ is the transformed input and $y$ is the output read) does not lead to an accepting state, in step~\ref{algoEo-Edit} we check the element corresponding to $(q,x')$ from the table and assign it to $y'$.
\end{remark}
\section{Application to SCCharts}
\label{sec:sccharts}
\newcommand{\tsf}[1]{\textsf{#1}}
\newcommand{\ie}{i.e.\xspace}
SCCharts is a Statechart dialect that has been designed for safety-critical systems and offers deterministic concurrency~\cite{vonHanxleden2014}.
We implemented the algorithm presented in Section~\ref{sec:algo} in an SCCharts compilation framework\footnote{\url{https://rtsys.informatik.uni-kiel.de/kieler}} according to the \emph{single-pass language-driven incremental compilation} approach~\cite{MotikaSvH14}.
Here, a safety automaton is automatically transformed into a synchronous enforcer using model-to-model transformations.
The generated enforcer has three concurrent regions, one for reading and editing the inputs,
one for invoking the tick function \texttt{ptick} with the edited inputs and a final one for processing and emitting the outputs.
The three components exactly match the steps of the algorithm presented in Section~\ref{sec:algo}.

\begin{figure}[htb]
\begin{centering}
\scalebox{0.2}{
\includegraphics[scale=1]{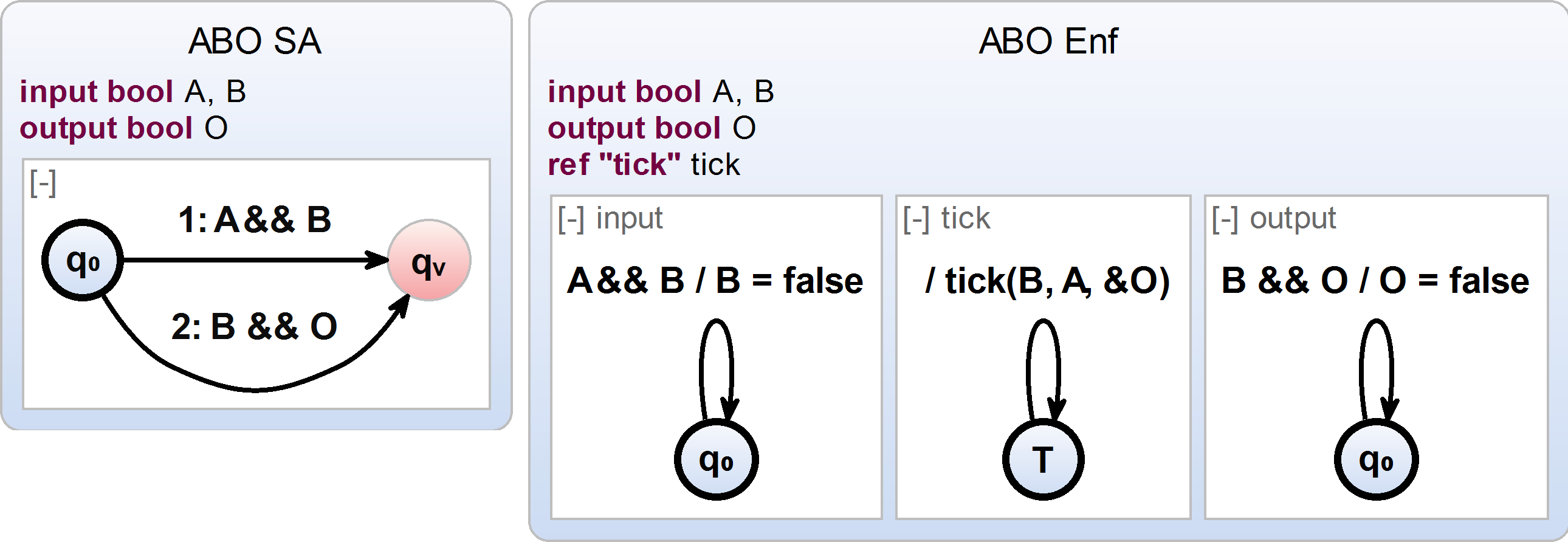}
}
\caption{Example safety automaton ABO in SCCharts (left) and its automatically generated enforcer (right).}
\label{fig:sccharts:abo}
\end{centering}
\end{figure}

Figure~\ref{fig:sccharts:abo} depicts the example safety automaton \tsf{ABO SA} in SCCharts and the automatically generated Enforcer \tsf{ABO Enf}.
In this example, \tsf{A} and \tsf{B} serve as input vector, whereas \tsf{O} is the only output. The automaton only has two states, the initial state $q_{0}$ and the violation state $q_{v}$. The safety property says that \tsf{A} and \tsf{B} and also \tsf{B} and \tsf{O} may not be present at the same time.
%
\begin{remark}
\label{rem:constructiveness}
In the ABO example, Figure~\ref{fig:sccharts:abo}, two regions (tick and output) concurrently write to a shared variable O. Usually, this would be considered a write-write race, leading either to non-determinism, e.g. in Java threads, or rejection at compile time due to non-causality~\cite{BenvenisteCEHLd03}, as in synchronous languages.
However, we can take advantage of the fixed execution sequence of the three regions during every tick, following Algorithm~\ref{algo:enf}.
First a transition in the input region is executed, followed by the tick region and finally the output region.
This approach, thus, follows the PRET-C~\cite{andalam14} semantics, which is causal by construction.
\end{remark}
\newcommand{\mn}[1]{\textsf{#1}}
In order to evaluate this implementation we used a series of models with increasing sizes as can be seen in Table~\ref{tableEval}.
To generate the mean values we simulated every model 5 times with each run consisting of 1000 ticks.
As inputs for each model, a random environment was created.
The whole setup was executed for two cases. Firstly, the plain model was simulated within its environment.
Secondly, the same environment was used to simulate the model again with an enforcer in between.
The number of enforced properties (entry ``$\#$ Properties'' in Table~\ref{tableEval}) range from 0 to 3 including properties that enforce inputs and also outputs (bi-directional properties).
All experiments were conducted on an embedded system equipped with an 1 GHz ARM Cortex-A7 Dual-Core.
Depending on the model size and the number of properties enforced, we see an increase of mean execution time between 12\%-38\% when simulating with an enforcer.
Due to the netlist-based code generation of KIELER, there is a constant overhead because of the tick function call. Therefore, the overhead decreases percentage-wise with increasing model size.
The \mn{Null} model test measures the overhead of this black-box call with an enforcer with 0 safety properties.
We observe a constant overhead of 0.1$\mu$s here.
%
%
\begin{table*}[tb]
\centering
\begin{tabular}{|c|c|c|c|c|c|c|c|}
\hline
\multicolumn{1}{|c|}{Examples\footnotemark} & \multicolumn{1}{c|}{Tick (LoC)} &\multicolumn{1}{c|}{$\#$ Properties} &\multicolumn{1}{c|}{Enf. (LoC)} & \multicolumn{1}{c|}{Time ($\mu$s)} & \multicolumn{1}{c|}{Time w/ Enf.  ($\mu$s)} & \multicolumn{1}{c|}{Incr. ($\%$)} \\
\hline
\mn{Null}  & 0 & 0 & 0 & 0.654 & 0.752  & 14.98 \\
\hline
\mn{ABRO} & 23 & 1 & 21 & 1.208 & 1.565  & 29.55\\
\hline
\mn{ABO}  & 28 & 1 & 21 & 0.998 & 1.368  & 37.10\\
\hline
\mn{Reactor} & 32 & 2 & 32 & 1.587 & 2.137 & 34.61\\
\hline
\mn{Faulty Heart Model} & 43 & 2 & 40 & 1.346 & 1.869 & 38.85\\
\hline
\mn{Simple Heart Model} & 76 & 2 & 40 & 2.175 & 2.825  & 29.86\\
\hline
\mn{Traffic Light} & 171 & 3 & 41 & 4.039  & 4.707 & 16.53\\
\hline
\mn{Pacemaker} & 271 & 2 & 35 & 7.302 & 8.318 & 13.91\\
\hline
\hline
\mn{FHM + Pacemaker} & 314 & 2 & 35 & 9.195 & 10.306 & 12.08 \\
\hline
\end{tabular}
\caption{Evaluation results.}
\label{tableEval}
\end{table*}
\footnotetext[2]{
\mn{ABRO} from~\cite{Berry00b},
\mn{ABO} from~\cite{vonHanxleden2014},
\mn{Reactor} from~\cite{TraulsenAvH11},
\mn{Simple Heart Model} and \mn{Pacemaker} are remodeled SCCharts variants from~\cite{JiangPMAM12},
\mn{Faulty Heart Model} is a variant of the \mn{Simple Heart Model} with deliberately flawed pulse signals,
\mn{Traffic Light} from~\cite{MotikaFvHL12} (remodeled from Ptolemy Traffic Light)
}

As a concrete case study, we selected a pacemaker based on~\cite{JiangPMAM12}, which has been implemented in SCCharts.
As second experiment we ran the \mn{Faulty Heart Model} together with the \mn{Pacemaker}.
The results of the close-loop simulation can be seen in the last row of Table~\ref{tableEval}.
Here, the \mn{Faulty Heart Model} serves as environment for the \mn{Pacemaker} and generates flawed pulse signals for the heart.
We added an enforcer to the pacemaker to make sure that atrial and ventricular signals cannot occur simultaneously,
which results in editing the input vector, and also that the pacemaker does not emit pace signals for both in return, which
results in editing the output vector.
We observe a mean overhead of 12\% when using the enforcer.
%
\section{Related Work}
\label{sec:rw}
Synthesizing enforcers from properties is an active area of research.
According to how an enforcer is allowed to correct the input sequence, several RE models have been proposed.
Security automata proposed by Schneider~\cite{enforceablesecpol} focus on enforcement of safety properties, where the enforcer blocks the execution when it recognizes a sequence of actions that doses not satisfy the desired property.
Edit automata~\cite{RuntimeNonSafety} allows the enforcer to correct the input sequence by suppressing and (or) inserting events, and the RE mechanisms proposed in~\cite{FalconeMFR11,FMSD} allows buffering events and releasing them upon observing a sequence that satisfies the desired property.
Recently, compositionality of enforcers has been studied in~\cite{cre2016}. Given a set of properties over the same alphabet, the problem studied in~\cite{cre2016} addresses whether it is possible to synthesize multiple enforcers, one for each property, and whether composing enforcers (in series or in parallel) can enforce all the properties.
Moreover, the enforcement framework in~\cite{cre2016} allows to buffer (delay) events.
These approaches focus on uni-directional RE.

Mandatory Result Automata (MRAs)~\cite{DolzhenkoLR15} extended edit-automata~\cite{RuntimeNonSafety}, by considering bi-directional runtime enforcement.
Compared to the other RE frameworks such as~\cite{enforceablesecpol,FalconeMFR11,RuntimeNonSafety,FMSD}, in MRA the focus is on handling communication between two parties.
However none of the above approaches are suitable for reactive systems since halting the program and delaying actions is not suitable.
This is because for reactive systems the enforcer has to react instantaneously.

Our work is closely related to~\cite{BloemKKW15}, which introduces a framework to synthesize enforcers for reactive systems, called as \emph{shields},
from a set of safety properties.
In our work, we restrict to prefix-closed safety properties. The approach in~\cite{BloemKKW15} seems to consider more that prefix-closed properties (where properties are expressed as automata), but not all regular properties.
Also, the approach in~\cite{BloemKKW15} has the notion of k-stabilization where the shield allows to deviate from the property for $k$ consecutive steps whenever a property violation is unavoidable.
If a second violation occurs within $k$ steps, then the shield enters into a  \textit{fail-safe} mode, where it ensures only correctness.
So, if two or more errors occur within k-steps, then the shield may generate outputs arbitrarily to satisfy the property being monitored by ignoring outputs from the system being monitored.
In our approach, if the input given to the enforcer satisfies the property, then the enforcer does not modify any event.
In case if a violation is noticed upon some event, the enforcer corrects it (to avoid violation), and continues to minimize deviation also for the future input events depending on the state of the enforcer and the received input event.
Moreover, in~\cite{BloemKKW15}, the shield is uni-directional, where it observes inputs from the environment and outputs from the system (program), and transforms erroneous outputs.
In our work, we consider bi-directional enforcement, as explained and illustrated in Fig.~\ref{fig:intro-context}.

Note that when we consider safety-critical embedded systems such as medical devices and automotive systems, it is also utmost important to monitor and transform ``illegal'' inputs, before they are fed to the program.
For instance, suppose that there are multiple sensors, and their values are inputs from the environment to the enforcer.
Some sensors may fail or may be attacked by some intruder. Unlike~\cite{BloemKKW15}, which ignores inconsistent inputs,
our work is able to deal with both the inputs (from the environment) and the outputs (from the synchronous program) simultaneously
during each reaction.
\section{Conclusions}
Synchronous observers are used to express safety properties for synchronous programs, which may be verified either statically or during runtime.
This paper extends observers by proposing the concept of runtime enforcers for synchronous programs.
The property to be enforced is modeled as a safety automaton, which is syntactically like an observer (expressed
as an automaton with a single violation state) referring to both inputs and outputs of the synchronous program.
We formalise, for the first time, the runtime enforcement synthesis problem for synchronous reactive systems.
We define enforceability conditions, provide an algorithm, and prove its correctness.
The synthesised enforcer interacts with a black-box synchronous program and its adjoining environment to
ensure that the property in question holds during program execution.
We have implemented the proposed enforcer synthesis algorithm for the SCCharts synchronous language.
We highlight the applicability of the proposed approach by enforcing policies over a synchronous pacemaker model.
In the near future, we will consider several extensions, including enforcement with valued
inputs and outputs (valued signals), non-safety properties, and
distributed enforcement.
%
\bibliographystyle{abbrv}
\bibliography{biblioShort}

\begin{thebibliography}{10}

\bibitem{andalam14}
S.~Andalam, P.~S. Roop, A.~Girault, and C.~Traulsen.
\newblock A predictable framework for safety-critical embedded systems.
\newblock {\em IEEE Transactions on Computers}, 63(7):1600--1612, 2014.

\bibitem{andre01}
C.~Andre, F.~Boulanger, and A.~Girault.
\newblock Software implementation of synchronous programs.
\newblock In {\em Application of Concurrency to System Design, 2001.
  Proceedings. 2001 International Conference on}, pages 133--142. IEEE, 2001.

\bibitem{BenvenisteCEHLd03}
A.~Benveniste, P.~Caspi, S.~Edwards, N.~Halbwachs, P.~Le~Guernic, and
  R.~de~Simone.
\newblock {The synchronous languages 12 years later}.
\newblock {\em Proceedings of the IEEE}, 91(1):64--83, Jan 2003.

\bibitem{Berry00b}
G.~Berry.
\newblock {\em The {Esterel} v5 Language Primer, Version v5\_91}.
\newblock Centre de Math{\'e}matiques Appliqu{\'e}es Ecole des Mines and INRIA,
  06565 Sophia-Antipolis, 2000.

\bibitem{BloemKKW15}
R.~Bloem, B.~K{\"{o}}nighofer, R.~K{\"{o}}nighofer, and C.~Wang.
\newblock Shield synthesis: Runtime enforcement for reactive systems.
\newblock In {\em {TACAS}}, volume 9035 of {\em LNCS}. Springer, 2015.

\bibitem{DolzhenkoLR15}
E.~Dolzhenko, J.~Ligatti, and S.~Reddy.
\newblock Modeling runtime enforcement with mandatory results automata.
\newblock {\em Int. J. Inf. Sec.}, 14(1):47--60, 2015.

\bibitem{FalconeHR13}
Y.~Falcone, K.~Havelund, and G.~Reger.
\newblock A tutorial on runtime verification.
\newblock In {\em Engineering Dependable Software Systems}, volume~34, pages
  141--175. {IOS} Press, 2013.

\bibitem{FalconeMFR11}
Y.~Falcone, L.~Mounier, J.-C. Fernandez, and J.-L. Richier.
\newblock Runtime enforcement monitors: composition, synthesis, and enforcement
  abilities.
\newblock {\em FMSD}, 38(3):223--262, 2011.

\bibitem{HalbwachsLR94}
N.~Halbwachs, F.~Lagnier, and P.~Raymond.
\newblock Synchronous observers and the verification of reactive systems.
\newblock In {\em Algebraic Methodology and Software Technology (AMAST’93)},
  pages 83--96. Springer, 1994.

\bibitem{JiangPMAM12}
Z.~Jiang, M.~Pajic, S.~Moarref, R.~Alur, and R.~Mangharam.
\newblock Modeling and verification of a dual chamber implantable pacemaker.
\newblock In {\em TACAS}, pages 188--203. Springer, 2012.

\bibitem{LeuckerS08jlap}
M.~Leucker and C.~Schallhart.
\newblock A brief account of runtime verification.
\newblock {\em Journal of Logic and Algebraic Programming}, 78(5):293--303,
  2009.

\bibitem{RuntimeNonSafety}
J.~Ligatti, L.~Bauer, and D.~Walker.
\newblock Run-time enforcement of nonsafety policies.
\newblock {\em ACM Trans. Inf. Syst. Secur.}, 12(3):19:1--19:41, Jan. 2009.

\bibitem{MotikaFvHL12}
C.~Motika, H.~Fuhrmann, R.~von Hanxleden, and E.~A. Lee.
\newblock Executing domain-specific models in {E}clipse.
\newblock Technical Report 1214, Christian-Albrechts-Universit{\"a}t zu Kiel,
  Department of Computer Science, October 2012.
\newblock ISSN 2192-6247.

\bibitem{MotikaSvH14}
C.~Motika, S.~Smyth, and R.~von Hanxleden.
\newblock {Compiling SCCharts--A} case-study on interactive model-based
  compilation.
\newblock In {\em ISoLA}, volume 8802 of {\em LNCS}, pages 443--462, Corfu,
  Greece, 2014.

\bibitem{FMSD}
S.~Pinisetty, Y.~Falcone, T.~J{\'e}ron, H.~Marchand, A.~Rollet, and
  O.~Nguena~Timo.
\newblock Runtime enforcement of timed properties revisited.
\newblock {\em FMSD}, 45(3):381--422, 2014.

\bibitem{cre2016}
S.~Pinisetty and S.~Tripakis.
\newblock Compositional runtime enforcement.
\newblock In {\em NASA Formal Methods Symposium, NFM 2016, Minneapolis, MN,
  USA}, pages 82--99. Springer, 2016.

\bibitem{RaymondNHW98}
P.~Raymond, X.~Nicollin, N.~Halbwachs, and D.~Weber.
\newblock Automatic testing of reactive systems.
\newblock In {\em Real-Time Systems Symposium}, pages 200--209. IEEE, 1998.

\bibitem{Rushby14}
J.~Rushby.
\newblock The versatile synchronous observer.
\newblock In {\em Specification, Algebra, and Software}, pages 110--128.
  Springer, 2014.

\bibitem{enforceablesecpol}
F.~B. Schneider.
\newblock Enforceable security policies.
\newblock {\em ACM Trans. Inf. Syst. Secur.}, 3(1):30--50, 2000.

\bibitem{TraulsenAvH11}
C.~Traulsen, T.~Amende, and R.~von Hanxleden.
\newblock Compiling {SyncCharts} to {Synchronous C}.
\newblock In {\em Proceedings of the Design, Automation and Test in Europe
  Conference (DATE'11)}, pages 563--566, Grenoble, France, March 2011. IEEE.

\bibitem{vonHanxleden2014}
R.~von Hanxleden, B.~Duderstadt, C.~Motika, S.~Smyth, M.~Mendler, J.~Aguado,
  S.~Mercer, and O.~O'Brien.
\newblock {SCCharts}: Sequentially constructive statecharts for safety-critical
  applications.
\newblock In {\em PLDI}, pages 372--383, NY, USA, 2014. ACM.

\end{thebibliography}
\appendix
\newpage
\section{Appendix: Proofs}
\label{sec:proofs}
\begin{proof}[of Theorem~\ref{rem:nonEnf}]
\label{proof:rem:nonEnf}
Let us recall Theorem~\ref{rem:nonEnf}.
Consider a property $\varphi$ defined as SA $\calA_\varphi=(Q, q_0, q_v, \Sigma, \rightarrow)$\footnote{Note that we consider that $\calA_\varphi$ is deterministic and complete, and $Q$ does not contain any (redundant) locations that are not reachable from $q_0$ in 1 or more steps.}.
Property $\varphi$ is enforceable iff the condition~(\ref{suff_cond_enforceability}) holds which is the following condition:
$\forall q \in Q, q \neq q_v \implies \exists (x,y) \in \Sigma: q \xrightarrow{(x,y)} q' \wedge q'\neq q_v $.

We prove that:
\begin{itemize}
\item \textit{Sufficient:} If condition~(\ref{suff_cond_enforceability}) holds then $\ef$ according to Definition~\ref{def-E-func-constraints} exists.

Due to condition~(\ref{suff_cond_enforceability}), whatever may be the current state $q\in Q\setminus\{q_v\}$ of the enforcer, there is at least one possibility to correct the event that it receives when in state $q$ (in case if the received event leads to $q_v$ from $q$).
That is, due to condition~(\ref{suff_cond_enforceability}), $\forall q \in Q\setminus\{q_v\}$, we know for sure that $\editIaut(q)$ will be non-empty, and $\forall q \in Q\setminus\{q_v\}, \forall x \in \editIaut(q): \editOaut(q,x)$ will be also non-empty.

For any property $\varphi$ defined as SA $\calA_\varphi$, the enforcement function $\efalgo$ (Definition~\ref{def-algo-ef}) is an enforcer for $\varphi$ which satisfies all the constraints according to Definition~\ref{def-E-func-constraints}.
Theorem~\ref{prop-constraints-algo} shows that for any property $\varphi$ (defined as SA $\calA_\varphi$) that satisfies the condition for enforceability~(\ref{suff_cond_enforceability}),  the enforcement function $\efalgo$ (Definition~\ref{def-algo-ef}) is an enforcer for $\varphi$, that is, it satisfies (\ref{eq:snd}), (\ref{eq:tr}), (\ref{eq:mono}), (\ref{eq:inst}), and (\ref{eq:ca}) constraints of Definition~\ref{def-E-func-constraints}.

\item \textit{Necessary:} If $\ef$ according to Definition~\ref{def-E-func-constraints} exists, then condition~(\ref{suff_cond_enforceability}) holds.

Suppose that an enforcer $\ef$ for $\varphi$ according to Definition~\ref{def-E-func-constraints} exists and assume that condition~(\ref{suff_cond_enforceability}) does not hold for $\calA_\varphi$.

Since condition~(\ref{suff_cond_enforceability}) does not hold, $\exists q\in Q\setminus\{q_v\}: \forall (x,y) \in \Sigma: q \xrightarrow{(x,y)} q_v$, i.e., there exists a location $q\in Q\setminus\{q_v\}$ such that all the outgoing transitions from $q$ go to $q_v$.

Since all the locations in $Q$ are reachable from $q_0$, $\exists \sigma\in\Sigma^*: q_0\xrightarrow{\sigma}q$, i.e., there certainly exists a word $\sigma\in\Sigma^*$ that leads to the problematic accepting location $q$ (which has all its outgoing transitions to $q_v$) from the initial location $q_0$.

If $\sigma$ is the input word to the enforcer, then due to constraint~(\ref{eq:tr}), it cannot edit any event in $\sigma$, and the enforcer produces $\sigma$ as output and reaches location $q$.
When in location $q$, upon receiving any event $(x,y)\in\Sigma$, the enforcer has no possibility to correct it, since every event in $\Sigma$ leads to $q_v$ from $q$ (i.e., since $\forall (x,y) \in \Sigma: q \xrightarrow{(x,y)} q_v$).

Thus, $\forall(x,y)\in \Sigma$, when the input word given to the enforcer is $\sigma\cdot(x,y)$, the enforcer cannot produce any event as output since $\editI()$ and $\editO()$ from location $q$ will be empty, violating constraints~(\ref{eq:inst}) and~(\ref{eq:ca}).
Thus, our assumption is false and condition~(\ref{suff_cond_enforceability}) holds for $\calA_\varphi$.
\end{itemize}
\end{proof}
\begin{proof}[of Theorem~\ref{prop-constraints-algo}]
\label{proof-prop-constraints-algo}
Let us recall the condition for enforceability:
A property $\varphi$ defined as SA $\calA_\varphi=(Q, q_0, q_v, \Sigma, \rightarrow)$ is enforceable iff
\[
\forall q \in Q, q \neq q_v \implies \exists (x,y) \in \Sigma: q \xrightarrow{(x,y)} q' \wedge q'\neq q_v.
\]

Let us also recall the definition of function $\efalgo:\Sigma^*\to\Sigma^*$ (Definition~\ref{def-algo-ef}).
Let $\sigma = (x_1, y_1) \cdots (x_k,y_k)\in\Sigma^*$ be a word received by Algorithm~\ref{algo:enf}.
Then we let $\efalgo(\sigma)=(x_1', y_1') \cdots (x_k',y_k')$, where $(x_t',y_t')$ is the pair of events output by Algorithm~\ref{algo:enf} in Step~\ref{algoEo-release},
for $t=1,\ldots,k$.

Note that the input automaton $\calA_{\varphi_I}=(Q, q_{0}, q_{v}, \Sigma_I, \rightarrow_I)$ is obtained from $\calA_{\varphi}$ by projecting on inputs (See Definition~\ref{def:inp:prop:proj:def}, Section~\ref{sec:prelim:re}).

We shall prove that given any safety property $\varphi$ defined as SA $\calA_\varphi$ that satisfies condition~(\ref{suff_cond_enforceability}),
the function $\efalgo$ is an enforcer for $\varphi$, that is, it satisfies (\ref{eq:snd}), (\ref{eq:tr}), (\ref{eq:mono}), (\ref{eq:inst}), and (\ref{eq:ca}) constraints of Definition~\ref{def-E-func-constraints}.

Let us prove this theorem using induction on the length of the input sequence $\sigma\in\Sigma^*$ (which also corresponds to the number of ticks/iterations of Algorithm~\ref{algo:enf}).

\textit{Induction basis.}
Theorem~\ref{prop-constraints-algo} holds trivially for $\sigma=\epsilon$ since the algorithm will not release any input-output event as output and thus $\efalgo(\epsilon)= \epsilon$.

\textit{Induction step.}
Assume that for every $\sigma = (x_1, y_1) \cdots (x_k,y_k) \in \Sigma^*$ of some length $k \in \bbn$, let $\efalgo(\sigma) =(x'_1, y'_1) \cdots (x'_k,y'_k) \in \Sigma^*$, for $t=1,\ldots,k$, and Theorem~\ref{prop-constraints-algo} holds for $\sigma$, i.e., $\efalgo(\sigma)$ satisfies the (\ref{eq:snd}), (\ref{eq:tr}), (\ref{eq:mono}), (\ref{eq:inst}), and (\ref{eq:ca}) constraints.
Let $q \in Q\setminus \{q_v\}$ be the current state of both the automata $\calA_\varphi$ and $\calA_{\varphi_I}$ after processing input $\sigma$ of length $k$, i.e., $q$ corresponds to the state that we reach upon $\efalgo(\sigma)$ in $\calA_\varphi$, and the state that we reach in the automaton $\calA_{\varphi_I}$ upon $\efalgo(\sigma)_I$.
Note that the current state $q$ in Algorithm~\ref{algo:enf} can never be $q_v$ ($q$ is initialized to $q_0$ and it is updated in step~\ref{algo-stateUpdate} to a state $q'\in Q\setminus\{q_v\}$).

We now prove that for any event $(x_{k+1},y_{k+1})\in\Sigma$, Theorem~\ref{prop-constraints-algo} holds for $\sigma\cdot(x_{k+1},y_{k+1})$, where $x_{k+1}\in\Sigma_I$ is the input event read by Algorithm~\ref{algo:enf}, and $y_{k+1} \in \Sigma_O$ is the output event read by Algorithm~\ref{algo:enf} in $k+1^{th}$ iteration (i.e., when $t= k+1$).
We have the following two possible cases based on whether there is a transition in the automaton $\calA_\varphi$ from the current state $q$ upon $(x_{k+1},y_{k+1})$ to an accepting state.

\begin{itemize}
\item {$\exists q'\in Q: q\xrightarrow{(x_{k+1},y_{k+1})} q' \wedge q' \neq q_{v}$.}

In Algorithm~\ref{algo:enf}, the condition tested in step~\ref{algoEi-test} will evaluate to true
since from Lemma~\ref{lem:inputProp}, in $\calA_{\varphi_I}$ we will have $\exists q' \in Q:  q \xrightarrow{x_{k+1}}_I q' \wedge q' \neq q_{v}$,
and thus $x'_{k+1} = x_{k+1}$.

Also, the condition tested in step~\ref{algoEo-test} will evaluate to true in this case since $\exists q'\in Q: q\xrightarrow{(x_{k+1},y_{k+1})} q' \wedge q' \neq q_{v}$,
and thus  $y'_{k+1} = y_{k+1}$.
At the end of the $k+1^{th}$ iteration, the input-output event released as output by the algorithm in step~\ref{algoEo-release} is $(x_{k+1},y_{k+1})$.
The output of the algorithm after completing the $k+1^{th}$ iteration is $\efalgo(\sigma\cdot(x_{k+1},y_{k+1}))= \efalgo(\sigma)\cdot(x_{k+1},y_{k+1})$.

Regarding constraint~(\ref{eq:snd}), in this case, what has been already released as output by the algorithm earlier before reading event $(x_{k+1},y_{k+1})$ (i.e., $\efalgo(\sigma)$) followed by the new input-output event released as output $(x_{k+1},y_{k+1})$ satisfies the property $\varphi$, and thus constraint~(\ref{eq:snd}) holds.

Regarding constraint~(\ref{eq:mono}), it holds since $\sigma \pref \sigma \cdot (x_{k+1},y_{k+1})$ and also $\efalgo(\sigma) \pref \efalgo(\sigma)\cdot(x_{k+1},y_{k+1})$.

Regarding constraint~(\ref{eq:inst}) from the induction hypothesis, we have for $\sigma$ of some length $k$, $|\sigma|=|\ef(\sigma)|$.
We also have $\efalgo(\sigma \cdot (x_{k+1},y_{k+1})) = \efalgo(\sigma)\cdot(x_{k+1},y_{k+1})$.
Thus, $|\sigma\cdot (x_{k+1},y_{k+1})| = |\efalgo(\sigma \cdot (x_{k+1},y_{k+1}))| = k+1$, and constraint~(\ref{eq:inst}) holds.

Constraint~(\ref{eq:tr}) holds in this case since the output of the enforcer before reading $(x_{k+1},y_{k+1})$ i.e., $\efalgo(\sigma)$ followed by the new input-output event read $(x_{k+1},y_{k+1})$ satisfies the property $\varphi$ and we already saw that the output event released by the algorithm after reading $(x_{k+1},y_{k+1})$ is $\efalgo(\sigma)\cdot(x_{k+1},y_{k+1})$.

Regarding constraint~(\ref{eq:ca}), in this case from the induction hypothesis, from the definitions of $\editIaut$ and $\editOaut$ we have $x_{k+1} \in \editIaut(q)$, and also $y_{k+1} \in \editOaut(q, x_{k+1})$.

Theorem~\ref{prop-constraints-algo} thus holds for $\sigma\cdot(x_{k+1},y_{k+1})$ in this case.

\item {$\nexists q'\in Q: q\xrightarrow{(x_{k+1},y_{k+1})} q' \wedge q' \neq q_{v}$.}

In this case, we have two sub-cases, based on whether $\exists q' \in Q: q \xrightarrow{x_{k+1}}_I q' \wedge q' \neq q_{v}$ in $\calA_{\varphi_I}$.

\begin{itemize}
\item $\exists q' \in Q: q \xrightarrow{x_{k+1}}_I q' \wedge q' \neq q_{v}$.

In Algorithm~\ref{algo:enf}, the condition tested in step~\ref{algoEi-test} will evaluate to true and thus $x'_{k+1} = x_{k+1}$.

In this case, the condition tested in step~\ref{algoEo-test} will evaluate to false since $\nexists q'\in Q: q\xrightarrow{(x_{k+1},y_{k+1})} q' \wedge q' \neq q_{v}$.
$y'_{k+1}$ will thus be an element belonging to the set $\editOaut(q,x_{k+1})$ if $\editOaut(q,x_{k+1})$ is non-empty.
It is important to notice that $\editOaut(q,x_{k+1})$ will be non-empty in this case since we know for sure that $\exists y'_{k+1} \in \Sigma_O, q'\in Q: q\xrightarrow{(x_{k+1},y'_{k+1})} q' \wedge q' \neq q_{v}$ (from the condition for enforceability~(\ref{suff_cond_enforceability}), hypothesis ($q \neq q_v$), definition of $\editOaut$, and Lemma~\ref{lem:inputProp}).
Thus $y'_{k+1}$ is an element belonging to $\editOaut(q,x_{k+1})$.
The output of the algorithm after completing the $k+1^{th}$ iteration is $\efalgo(\sigma\cdot(x_{k+1},y_{k+1})) =\efalgo(\sigma)\cdot(x_{k+1},y'_{k+1})$.

Regarding constraint~(\ref{eq:snd}), from the definition of $\editOaut$, we know that $\efalgo(\sigma)$ followed by the new input-output event released as output $(x_{k+1},y'_{k+1})$ satisfies property $\varphi$, and thus constraint~(\ref{eq:snd}) holds.

The reasoning for constraints~(\ref{eq:mono}) and~(\ref{eq:inst}) are similar to the previous cases since we saw that Algorithm~\ref{algo:enf} releases a new event $(x_{k+1},y'_{k+1})$ as output after reading event $(x_{k+1},y_{k+1})$ after completing $k+1^{th}$ iteration.

Constraint~(\ref{eq:tr}) holds trivially in this case since $\efalgo(\sigma)\cdot(x_{k+1},y_{k+1})\not\models\varphi$.

Regarding constraint~(\ref{eq:ca}), in this case from the induction hypothesis, from the definitions of $\editIaut$ we have $x_{k+1} \in \editIaut(q)$,
and we already discussed that $\editOaut(q,x_{k+1})$ will be non-empty and thus constraint~(\ref{eq:ca}) holds in this case.

\item $\nexists q' \in Q: q \xrightarrow{x_{k+1}}_I q'\wedge q' \neq q_{v}$.

In Algorithm~\ref{algo:enf}, the condition tested in step~\ref{algoEi-test} will evaluate to false in this case.
It is important to notice that $\editIaut(q)$ will be non-empty since from the condition for enforceability and Lemma~\ref{lem:inputProp}, we know for sure that $\exists x'\in\Sigma_I, q'\in Q: q\xrightarrow{x'_{k+1}} q' \wedge q' \neq q_{v}$ in the automaton $\calA_{\varphi_I}$.
Thus, $x'_{k+1}$ will be an element belonging to $\editIaut(q)$.

We have two sub-cases based on whether $\exists q' \in Q: q\xrightarrow{(x'_{k+1},y_{k+1})} q' \wedge q' \neq q_{v}$ or not.

\begin{itemize}
\item $\exists q' \in Q: q\xrightarrow{(x'_{k+1},y_{k+1})} q' \wedge q' \neq q_{v}$.

In Algorithm~\ref{algo:enf}, the condition tested in step~\ref{algoEo-test} will evaluate to true in this case.
Thus, $y'_{k+1} = y_{k+1}$ in this case and the event released as output by the algorithm at the end of $k+1^{th}$ iteration is $(x'_{k+1}, y_{k+1})$.
We have $\efalgo(\sigma\cdot(x_{k+1}, y_{k+1})) = \efalgo(\sigma)\cdot(x'_{k+1}, y_{k+1})$.

Regarding constraint(~\ref{eq:snd}), from the condition of this case (i.e., $\exists q' \in Q: q\xrightarrow{(x'_{k+1},y_{k+1})} q' \wedge q' \neq q_{v}$), we know that $\efalgo(\sigma)$ followed by the new input-output event released as output $(x'_{k+1},y_{k+1})$ satisfies the property $\varphi$, and thus constraint~(\ref{eq:snd}) holds.

The reasoning for constraints~(\ref{eq:mono}) and~(\ref{eq:inst}) are similar to the previous cases since we saw that the algorithm releases a new event $(x'_{k+1},y_{k+1})$ as output after reading event $(x_{k+1},y_{k+1})$ at the end of $k+1^{th}$ iteration.

Constraint~(\ref{eq:tr}) holds trivially in this case since $\efalgo(\sigma)\cdot (x_{k+1},y_{k+1}) \not\models\varphi$.

Regarding constraint~(\ref{eq:ca}), we already discussed that $\editIaut(q)$ is non-empty and $x'_{k+1} \in \editIaut(q)$, and $y_{k+1} \in \editOaut(q, x'_{k+1})$ from the condition of this case and definitions of $\editIaut$ and $\editOaut$.
\item $\nexists q' \in Q: q\xrightarrow{(x'_{k+1},y_{k+1})} q' \wedge q' \neq q_{v}$.

In the algorithm, the condition tested in step~\ref{algoEo-test} will evaluate to false in this case.

$y'_{k+1}$ will thus be an element belonging to the set $\editOaut(q,x'_{k+1})$ if $\editOaut(q,x'_{k+1})$ is non-empty.
Note that $\editOaut(q,x'_{k+1})$ will be non-empty in this case since we know for sure that $\exists y'_{k+1} \in \Sigma_O, q'\in Q: q\xrightarrow{(x'_{k+1},y'_{k+1})} q' \wedge q' \neq q_{v}$ (from the enforceability condition, definitions, and Lemma~\ref{lem:inputProp}).
Thus $y'_{k+1}$ is an element belonging to $\editOaut(q,x'_{k+1})$.
The output of the algorithm after completing the $k+1^{th}$ iteration is $\efalgo(\sigma\cdot(x_{k+1},y_{k+1}))=\efalgo(\sigma)\cdot(x'_{k+1},y'_{k+1})$ where $x'_{k+1}$ is an element belonging to $\editIaut(q_I)$ and $y'_{k+1}$ is an element belonging to $\editOaut(q, x'_{k+1})$.

Regarding constraint~(\ref{eq:snd}), from the definitions of $\editIaut$ and $\editOaut$, we know that $\ef(\sigma)\cdot(x'_{k+1},y'_{k+1})$ satisfies the property $\varphi$ and thus constraint~(\ref{eq:snd}) holds.

The reasoning for constraints~(\ref{eq:mono}) and~(\ref{eq:inst}) are similar to the previous cases since we saw that the algorithm releases a new event $(x'_{k+1},y'_{k+1})$ as output after reading event $(x_{k+1},y_{k+1})$.

Constraint~(\ref{eq:tr}) holds trivially in this case since $\ef(\sigma)\cdot (x_{k+1},y_{k+1}) \not\models\varphi$.

Regarding constraint~(\ref{eq:ca}), we already discussed that $\editIaut(q)$ is non-empty and $x'_{k+1} \in \editIaut(q)$, and $\editOaut(q, x'_{k+1})$ is also non-empty and $y'_{k+1} \in \editOaut(q, x'_{k+1})$ and thus constraint~(\ref{eq:ca}) holds.
\end{itemize}
\end{itemize}
Theorem~\ref{prop-constraints-algo} thus holds for $\sigma\cdot(x_{k+1},y_{k+1})$ in this case.
\end{itemize}
Thus Theorem~\ref{prop-constraints-algo} holds for $\sigma\cdot(x_{k+1},y_{k+1})$.
\end{proof}
\section{Appendix: Transformation of Non-Enforceable Properties}
\label{ap:trans:nonEnf}
Let us recall the discussion about non-enforceable properties in Section~\ref{sec:problemDef} (Example~\ref{eg:nonEnf}, and the condition for enforceability (\ref{suff_cond_enforceability})).
We also saw that some non-enforceable properties can be transformed in to enforceable properties (by excluding some behaviors from the given non-enforceable property) via an example discussed in Remark~\ref{rem:transfo:nonEnf}, and an algorithm for transformation of non-enforceable properties is also briefly discussed in Section~\ref{sec:problemDef} after Remark~\ref{rem:transfo:nonEnf}.

Let us now discuss in detail about an algorithm that takes a safety automaton $\calA_\varphi$ that does not satisfy the condition for enforceability (\ref{suff_cond_enforceability}) and checks whether $\calA_\varphi$ can be transformed into an enforceable property (by excluding some behaviors) or not.
If $\calA_\varphi$ can be transformed, then the algorithm returns the transformed safety automaton $\mathsf{sub}(\calA_\varphi)$.
The algorithm excludes only problematic paths (behaviors) from $\calA_\varphi$, and all good behaviors will be retained in $\mathsf{sub}(\calA_\varphi)$ (i.e., removal of behaviors is done minimally).

%
\newcommand{\return}{\mathsf{RETURN}}
\newcommand{\remove}{\mathsf{remove}}
\begin{algorithm}[ht]
\caption{$\mathsf{Transform NonEnf}$}
\label{algo:trans:non-enf}
{
\begin{algorithmic}[1]
   \STATE $sub(\calA_\varphi) \gets \calA_\varphi = (Q, q_0, q_v, \Sigma, \xrightarrow{})$
    \WHILE {$\exists q \in Q\setminus \{q_v\}: \forall(x,y)\in\Sigma,q\xrightarrow{(x,y)}q_v$}
    \FORALL {$q\in Q\setminus \{q_v\}$}
        \IF {$\forall(x,y)\in\Sigma, q\xrightarrow{(x,y)}q_v$}
            \IF{$q=q_0$}
                \STATE $\return$(NONE)
            \ELSE
             \STATE $\remove(q)$
            \ENDIF
        \ENDIF
        \ENDFOR
    \ENDWHILE
\STATE $\return(\mathsf{sub}(\calA_\varphi))$
\end{algorithmic}
}
\end{algorithm}
%
Algorithm~\ref{algo:trans:non-enf} takes an SA $\calA_\varphi= (Q, q_0, q_v, \Sigma, \xrightarrow{})$ (that does not satisfy the condition for enforceability) as input and returns an SA $\mathsf{sub}(\calA_\varphi)$ which is enforceable or it returns NONE if $\calA_\varphi$ cannot be transformed into an enforceable property.
$\mathsf{sub}(\calA_\varphi)$ is initialized with the $\calA_\varphi$.
Function $\remove$ takes a state $q\in Q\setminus \{q_v\}$ and merges it with $q_v$ (i.e., $q$ is removed from the set of states $Q$ and all the incoming transitions to $q$ go to $q_v$ instead).

The algorithm proceeds as follows:
The condition of the while loop tests whether there are any states in $\mathsf{sub}(\calA_\varphi)$ that have all its outgoing transitions to $q_v$.
If this condition evaluates to true, then each state $q$ in $Q\setminus \{q_v\}$ is checked (whether all the outgoing transitions from $q$ go to $q_v$).
If $q= q_0$ is such a state (i.e., all the outgoing transitions from $q_0$ go to $q_v$)), then the algorithm immediately returns ``NONE'' (i.e., that $\calA_\varphi$ cannot be transformed into an enforceable property). Otherwise, if $q$ is different from $q_0$ and if all the outgoing transitions from $q$ go to $q_v$, then state $q$ is removed and merged with $q_v$.
Finally, when there are no states in $Q\setminus\{q_v\}$ with all outgoing transitions to $q_v$, the while loop ends and the transformed automaton $\mathsf{sub}(\calA_\varphi)$ is returned.
%

Note that if the automaton that is given as input to the algorithm already satisfies the condition for enforceability (\ref{suff_cond_enforceability}), then the algorithm returns the same automaton (the while loop condition test will evaluate to false and thus is never executed).
\begin{example}
Let us now consider some examples.
Consider the example non-enforceable property discussed in Section~\ref{sec:problemDef}, presented in Figure~\ref{fig:prop1:nonEnf}.
If the property automaton in Figure~\ref{fig:prop1:nonEnf} is given as input to Algorithm~\ref{algo:trans:non-enf}, in the first iteration, the while condition test will evaluate to true since there is a state $q_1$, such that all the outgoing transitions from $q_1$ go to $q_v$ (i.e., $q_1\xrightarrow{\Sigma}q_v$).
Thus, in the first iteration, $q_1$ will be merged with $q_v$ (i.e., $q_1$ will be removed and all the incoming transitions to $q_1$ go to $q_v$).
Before the second iteration of the while loop starts, we will have only two locations $q_0$ and $q_v$ in the automaton, where all the transitions from location $q_0$ go to $q_v$.
In the second iteration the while condition will evaluate to true, and the algorithm returns NONE, since $q_0\xrightarrow{\Sigma}q_v$ and the initial locations also needs to be removed.
\end{example}
\begin{example}
We already discussed in Section~\ref{sec:problemDef} that the example non-enforceable presented in Figure~\ref{fig:nonenf:prop2} can be transformed to an enforceable property.
The automaton in Figure~\ref{fig:nonenf:prop2:trans} presents the transformed automaton returned by Algorithm~\ref{algo:trans:non-enf}.
\end{example}
%
\end{document}